\documentclass[american,aps,prxquantum,reprint,superscriptaddress, longbibliography,floatfix,nofootinbib]{revtex4-2}
\usepackage[T1]{fontenc}
\usepackage{inputenc}
\usepackage{color}
\usepackage{ragged2e}
\usepackage{centernot}
\usepackage{babel}
\usepackage{physics}
\usepackage{mathtools}
\usepackage{bbding}
\usepackage{pifont}
\usepackage{textcomp}
\usepackage{soul}
\usepackage{wasysym}
\usepackage{amsthm}
\usepackage{nicefrac}
\usepackage{wasysym}
\usepackage{dsfont}
\usepackage{amsmath}
\usepackage{amssymb}
\usepackage{subfig}
\usepackage{graphicx}
\usepackage{subcaption}
\usepackage{enumitem}
\definecolor{darkblue}{rgb}{0.0, 0.0, 0.55}
\usepackage[colorlinks=true,
            allcolors=blue]{hyperref}
\hypersetup{
     colorlinks   = true,
     linkcolor    = blue,
     citecolor    = blue,
     urlcolor     = blue,
     }
\usepackage{appendix}
\renewcommand{\selectlanguage}[1]{}
\makeatletter
\@ifundefined{textcolor}{}
{%
	\definecolor{BLACK}{gray}{0}
	\definecolor{WHITE}{gray}{1}
	\definecolor{RED}{rgb}{1,0,0}
	\definecolor{GREEN}{rgb}{0,1,0}
	\definecolor{BLUE}{rgb}{0,0,1}
	\definecolor{CYAN}{cmyk}{1,0,0,0}
	\definecolor{MAGENTA}{cmyk}{0,1,0,0}
	\definecolor{YELLOW}{cmyk}{0,0,1,0}
}
\theoremstyle{plain}

\theoremstyle{plain}

\ifx\proof\undefined
\newenvironment{proof}[1][\protect\proofname]{\par
	\normalfont\topsep6\p@\@plus6\p@\relax
	\trivlist
	\itemindent\parindent
	\item[\hskip\labelsep
	\scshape
	#1]\ignorespaces
}{%
	\endtrivlist\@endpefalse
}
\theoremstyle{remark} 
\newtheorem*{remark}{Remark} 
\providecommand{\proofname}{Proof}
\fi
\theoremstyle{plain}

\providecommand{\lemmaname}{Lemma}
\providecommand{\definitionname}{Definition}
\providecommand{\propositionname}{Proposition}

\usepackage{babel}
\usepackage{txfonts}
\usepackage{colortbl}\definecolor{myurlcolor}{rgb}{0,0,0.7}

\def\ket#1{| #1 \rangle}
\def\bra#1{\langle  #1 |}

\def\proj#1{| #1 \rangle\!\langle #1 |}
\usepackage[normalem]{ulem}
\usepackage{tikz}
\usetikzlibrary{circuits.ee.IEC}
\usetikzlibrary{positioning, shapes, arrows.meta, fit, calc}
\usetikzlibrary{arrows.meta,decorations.pathmorphing, patterns}

\newcommand{\haH}

\definecolor{orange}{RGB}{255,127,0}

\def\proj#1{| #1 \rangle\!\langle #1 |}

\renewcommand{\leq}{\leqslant}
\renewcommand{\geq}{\geqslant}
\renewcommand{\le}{\leqslant}
\renewcommand{\ge}{\geqslant}

\begin{document}
\title{Quantum SWITCH-induced non-Markovianity is not entirely quantum}

\author{Rajeev Gangwar}
\email{raju.gangwar420@gmail.com}
\affiliation{Technion - Israel Institute of Technology, Faculty of Mathematics, Haifa 3200003, Israel}

\author{Ananda G. Maity}
\email{anandamaity289@gmail.com}
\affiliation{School of Physical Sciences, Indian Institute of Technology Goa, Ponda 403401, Goa, India}

\begin{abstract}
Indefinite causal order extends quantum information processing beyond fixed causal structures, with the quantum SWITCH serving as its canonical realization. By coherently superposing different orders of quantum channels, the quantum SWITCH has been shown to provide operational advantages in communication, computation, metrology, and related tasks. Despite these advances, the physical resources responsible for these advantages remains unclear. Recent studies have further revealed that the quantum SWITCH can generate memory effects, manifested as non-Markovian information backflow. In this work, we examine the origin of such memory and determine whether they reflect genuine (quantum) non-Markovianity or instead arises from classical origin. To this end, we analyze two representative scenarios: one based on discrete-time evolution and another formulated through dynamical maps in open quantum systems. We show that the memory effects generated by the quantum SWITCH are not genuinely quantum non-Markovian, thereby prompting a re-examination of the source of quantum advantage in indefinite causal order frameworks.
\end{abstract}

\maketitle

\section{Introduction}\label{S1}
The foundations of modern information and communication theory were established by the pioneering work of Claude Shannon in 1948~\cite{Shannon_48}. While Shannon's framework was rooted in classical physics, the development of quantum mechanics led to the emergence of quantum information theory, where information is encoded, transmitted, and processed using quantum systems~\cite{nielsen_chuang_2010}. By exploiting fundamental quantum mechanical concepts such as superposition and entanglement, quantum information processing enables communication and computational protocols that surpass the fundamental capabilities achievable within the classical framework. These conceptual and technological advances formally led to the development of quantum Shannon theory, which extends the principles of classical information theory to the quantum domain and provides a unified framework for the storage, transmission, and manipulation of quantum information~\cite{nielsen_chuang_2010,Wilde13}.

Yet in conventional quantum Shannon theory, the configuration of quantum channels, namely their arrangement in space and time, has traditionally been assumed to follow a well-defined and fixed causal structure. However, quantum theory fundamentally permits scenarios in which the order of operations itself can exist in quantum superposition, giving rise to the framework of indefinite causal order, where the temporal arrangement of quantum processes is no longer predetermined~\cite{Hardy05,hardy2007towards}. In such settings, information carriers evolve through a coherent superposition of alternative operational sequences rather than a single definite ordering~\cite{Chiribella_13,Chiribella_12,Oreshkov_12}. 

An important primitive enabling indefinite causal order is the quantum SWITCH, which employs an auxiliary control system to coherently regulate the sequence of quantum operations~\cite{Chiribella_13,Chiribella_12}. Let $\mathcal{N}_1$ and $\mathcal{N}_2$ denote two quantum channels acting on a target system. The ordering of these channels is conditioned on the state of a control qubit. Specifically, if the control qubit is prepared in $\ket{0}_c$, the resulting evolution corresponds to the ordered composition $\mathcal{N}_2 \circ \mathcal{N}_1$, whereas preparation in $\ket{1}_c$ yields the reversed composition $\mathcal{N}_1 \circ \mathcal{N}_2$. When the control qubit is initialized in a coherent superposition state, such as $\ket{+}_c = (\ket{0}_c+\ket{1}_c)/\sqrt{2}$, the two causal orders become coherently superposed, realizing a process with indefinite causal order. Till then Indefinite causal order has emerged as a powerful operational resource in quantum information science. It enables enhanced discrimination and characterization of quantum channels, improves communication~\cite{Ebler_18,Chiribella_21,Bhattacharya_21} and computational performance~\cite{Araujo14}, reduces communication complexity~\cite{Guerin16}, and offers advantages in precision measurements~\cite{Zhao_20} and quantum thermodynamic processes~\cite{Tamal_20,Vedral_20,Mukhopadhyay19,Liu_23}. A growing body of experimental realizations has further confirmed these predicted benefits, demonstrating that indefinite causal structures can be harnessed as practical resources beyond conventional causally ordered protocols~\cite{Procopio15,Rubino17,Goswami18}.

Although indefinite causal order and the quantum SWITCH have demonstrated remarkable advantages across diverse quantum information-processing tasks, the physical origin of these advantages remains only partially understood. In particular, a clear identification of the fundamental resources responsible for the performance enhancement provided by the quantum SWITCH is still an open problem. Recent studies have emphasized that preserving quantum coherence in the control qubit, both during state preparation and measurement, is essential for observing operational advantages arising from indefinite causal order~\cite{Anand_25}. Further investigations have revealed that the magnitude of these advantages can also depend sensitively on the Hilbert-space dimension of the control system, suggesting that structural properties of the control degree of freedom play a crucial role in determining the performance of the protocol~\cite{Mukherjee24}. From a complementary dynamical perspective, quantum memory effects, especially non-Markovian dynamics, have likewise been proposed as a potential resource underlying the operation of the quantum SWITCH. In particular, it has been shown that the dynamics generated by the quantum SWITCH can exhibit signatures of non-Markovianity, indicating that information backflow may contribute significantly to the observed operational advantages~\cite{Maity_24,Anand_25}. 

Motivated by these developments, an important open question arises: whether the performance enhancement associated with the quantum SWITCH is genuinely rooted in intrinsically quantum memory effects, namely non-Markovianity, or whether it can be attributed merely to classical forms of memory. In this work, we undertake a systematic analysis of the memory underlying the advantages attributed to the quantum SWITCH. Our results indicate that the induced memory effects do not correspond to genuinely quantum non-Markovianity. This finding calls into question the physical origin of the reported advantages of indefinite causal order and invites a more careful reassessment of its role in quantum information processing.

The structure of the paper is as follows. In Sec.~\ref{S2}, we provide a concise overview of the quantum SWITCH and the notion of non-Markovian memory, including criteria for identifying genuinely quantum non-Markovian effects. Sec.~\ref{S3} contains our main results, where we analyze SWITCH-induced memory in two representative scenarios and show that the observed non-Markovian features are not genuinely quantum. Finally, Sec.~\ref{S4} concludes the paper by summarizing our results and highlighting avenues for future research.

\section{Preliminaries}\label{S2}
In this section, we briefly review the formalism of the quantum SWITCH and standard notions of non-Markovianity, including the distinction between genuine and non-genuine forms. In particular, here, we introduce the notations and framework needed for the subsequent analysis. Throughout the manuscript, we adopt conventional quantum information notation unless stated otherwise.

\subsection{Quantum SWITCH} \label{subsec:QS}
\begin{figure}[t!]
\centering
\includegraphics[width=8.5cm]{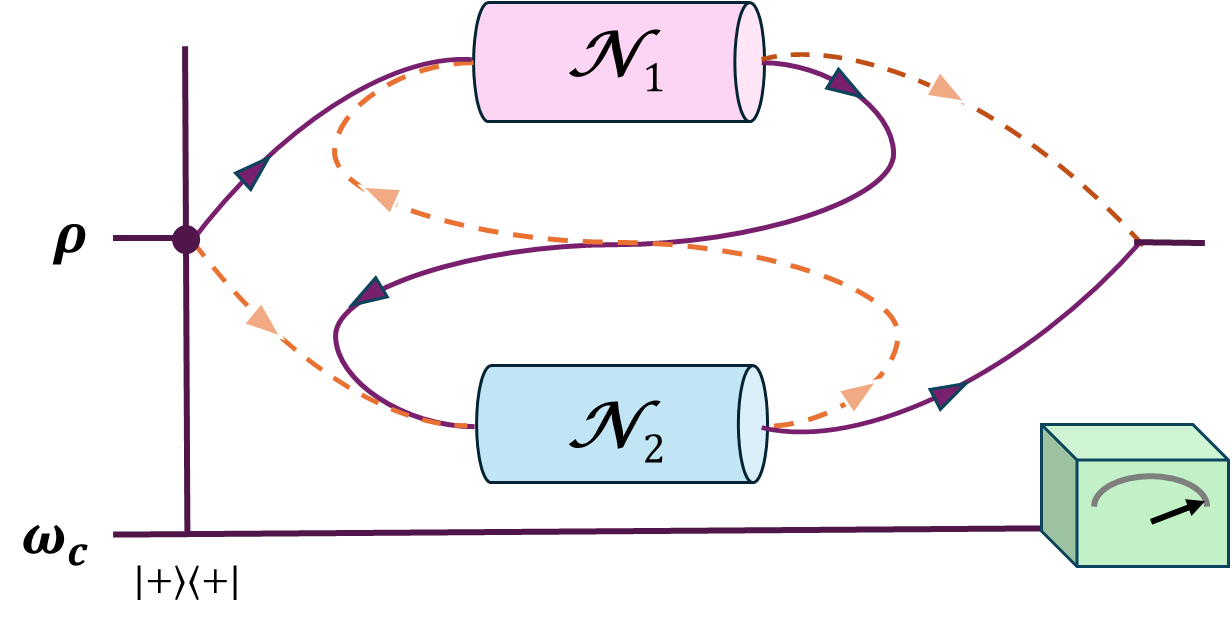}
        \caption{\justifying Schematic illustration of the quantum SWITCH. The control qubit is initialized in the state $\omega_c=\ket{+}\bra{+}$, thereby coherently superposing the two possible channel orders: $\mathcal{N}_1$ followed by $\mathcal{N}_2$, and $\mathcal{N}_2$ followed by $\mathcal{N}_1$. The control qubit is subsequently measured in the superposition basis.}
    \label{Fig1}
\end{figure}

A quantum channel $\mathcal{N}$ is a completely positive and trace-preserving (CPTP) map that transforms a valid quantum state (density operator $\rho$) defined on an input Hilbert space $\mathcal{H}_A$, i.e., $\rho \in \mathcal{D}(\mathcal{H}_A)$, into another valid quantum state in an output Hilbert space $\mathcal{H}_B$, i.e., $\mathcal{N}(\rho) \in \mathcal{D}(\mathcal{H}_B)$ where $\mathcal{D}(\mathcal{H}_i)$ denotes the set of density operators in Hilbert space $\mathcal{H}_i$. The action of a quantum channel admits a Kraus (operator-sum) representation given by $\mathcal{N}(\rho) = \sum_i K_i \rho K_i^{\dagger}$ where ${K_i}$'s are the Kraus operators, satisfying the completeness relation $\sum_i K_i^\dagger K_i = \mathbb{I}$, which guarantees that the channel is trace preserving. For two quantum channels, denoted by $\mathcal{N}_1$ and $\mathcal{N}_2$, their joint action may be implemented either in parallel or in sequence. The parallel application of the channels is described by the tensor product map $\mathcal{N}_1 \otimes \mathcal{N}_2$. In contrast, sequential compositions admit two possible configurations, namely $\mathcal{N}_1 \circ \mathcal{N}_2$ and $\mathcal{N}_2 \circ \mathcal{N}_1$, corresponding to different causal orders. Within the framework of definite causal structure, only one of these orderings can be realized at a time. However, as discussed earlier, indefinite causal order can be introduced by employing an additional control system, represented by a control qubit with state $\omega_c$, which coherently determines the order in which the channels act. Specifically, when the control qubit is prepared in the state $\omega_c=\ket{0}_c\bra{0}$, the channels are applied in the order $\mathcal{N}_1 \circ \mathcal{N}_2$, whereas preparation in $\omega_c=\ket{1}_c\bra{1}$ results in the reversed ordering $\mathcal{N}_2 \circ \mathcal{N}_1$. Let ${K_i^{(1)}}$ and ${K_j^{(2)}}$ denote the sets of Kraus operators corresponding to the channels $\mathcal{N}_1$ and $\mathcal{N}_2$, respectively. The general Kraus operators describing the quantum SWITCH are then given by \cite{Chiribella_12,Ebler_18,Chiribella_21}:
\begin{align}
    S_{i j}=K_{j}^{(2)} \circ K_{i}^{(1)} \otimes|0\rangle_{c}\left\langle 0\right|+K_{i}^{(1)} \circ K_{j}^{(2)} \otimes\left| 1\right\rangle_{c}\langle 1| .
\end{align}
The overall evolution of the joint system, consisting of the target state $\rho$ and the control qubit $\omega_c$, is therefore expressed as
$$
S\left(\mathcal{N}_1, \mathcal{N}_2 \right)\left(\rho \otimes \omega_{c}\right)=\sum_{i, j} S_{i j}\left(\rho \otimes \omega_{c}\right) S_{i j}^{\dagger}.
$$
Finally, the control qubit is measured in the coherent basis $\{\ket{+}\bra{+},\ket{-}\bra{-}\}$. Conditioned on each measurement outcome, the corresponding post-selected state of the target system is obtained as
  \begin{align}
     {}_c\bra{\pm}S(\mathcal{N}_1,\mathcal{N}_2)(\rho \otimes \omega_c) \ket{\pm}_c.
 \end{align}
The action of the quantum SWITCH, viewed as a coherent superposition of different channel orderings, is illustrated in Fig.\ref{Fig1}.

\subsection{Non-Markovian memory effect: A brief overview}

Non-Markovian dynamics have emerged as a promising route to quantum
advantages over conventional memoryless processes~\cite{breuer,rivas1,breuerN,alonso}, finding important use in long-distance quantum communication~\cite{task1,task3}, enhanced channel capacities~\cite{task2}, and quantum thermodynamic machines~\cite{task4}. A recurring difficulty, however, is separating the different origins of this memory. Classical non-Markovianity arises when past states influence the future through purely classical correlations~\cite{Buscemi_2025,Gangwar_2025}, whereas genuinely quantum non-Markovianity comes from system-environment correlations that have no classical description at all \cite{gangwar2026}. We briefly recall the standard tools used to detect non-Markovianity before addressing, in Sec.~\ref{sec:genuine}, the question of whether the observed memory effects are genuinely quantum or are merely classical.

One of the most common characterization of quantum Markovianity is based on CP-divisibility~\cite{rivas1}. A family of dynamical maps $\{\Lambda(t,t_0)\}_{t\ge t_0}$ is said to be Markovian if it can always be decomposed as $\Lambda(t,t_0)=\Lambda(t,s)\circ\Lambda(s,t_0)$ for all $t\ge s\ge t_0$, where the intermediate map $\Lambda(t,s)$ itself is completely positive. A breakdown of this condition signifies non-Markovian dynamics and is captured by the Rivas-Huelga-Plenio (RHP) measure
\cite{RHP},
\begin{align}
\mathcal{N}_{RHP}=\int_0^\infty g(t)\,dt ,\label{eq:RHP_measure}
\end{align}
where $g(t) = \lim_{ \Delta t\to 0} \frac{||(\mathbb{I} \otimes \Lambda (t+\Delta t,t)) \ket{\Phi}\bra{\Phi}||_{1}-1}{\Delta t}$. 
Here, $\Lambda (t+\Delta t,t)= (\mathbb{I}+ \Delta t\mathcal{L}_{t})$ corresponds to an infinitesimal dynamical map generated by a time-local Lindblad operator $\mathcal{L}_{t}$ and $\ket{\Phi}\bra{\Phi}$ denotes a maximally entangled state. For CP-divisible (Markovian) dynamics, $g(t)=0$ at all times, whereas $g(t)>0$ indicates a temporary breakdown of CP-divisibility. Eq.~\eqref{eq:RHP_measure} sums this contribution to quantify the degree of non-Markovianity.

An alternative (but more useful for our purposes) characterization of non-Markovianity was introduced by Luo, Fu, and Song (LFS)~\cite{Luo2012}. Rather than examining the divisibility of the dynamical map, this approach monitors the evolution of correlations between the system $Q$ and an isolated ancilla $R$, which is initially correlated with $Q$ but remains isolated from the environment and does not participate in the subsequent dynamics. The relevant quantity is the quantum mutual information (QMI), $I(R;Q)_t=S(\rho_R(t))+S(\rho_Q(t))-S(\rho_{RQ}(t))$ with $S(\cdot)$ being the von Neumann entropy. For Markovian dynamics, the data-processing inequality (DPI) guarantees that correlations between $R$ and $Q$ cannot increase, implying $\frac{d}{dt}I(R;Q)_t\le0$. Any temporary increase (or violation of DPI), therefore, signifies a backflow of information from the environment to the system. The corresponding LFS measure is defined as~\cite{Luo2012}
\begin{align}
\mathcal{N}_{QMI}=\max_{\rho_{RQ}(0)}\int_{\mu(t)>0}\mu(t)\,dt,
\qquad \mu(t):=\frac{d}{dt}I(R;Q)_t .
\end{align}
This is the criterion we will use throughout the paper. A closely related, and more refined version of this criterion
is followed by bringing the environment $E$ explicitly into the
picture. Using the chain rule for mutual information, we can write
$I(R;QE)=I(R;E|Q)+I(Q;R)$, where $I(R;E|Q)$ is the conditional mutual information (CMI). Since only $Q$ and $E$ undergo joint unitary evolution while $R$ remains untouched, the total correlation $I(R;QE)$ is conserved. Consequently, any increase in $I(R;Q)$ must be exactly balanced by a decrease in $I(R;E|Q)$. Because the CMI is always non-negative by strong subadditivity~\cite{Hayden_2004}, a decrease in $I(R;E|Q)$ represents a flow of information from the environment back to the system. This yields the equivalent non-Markovianity criterion $\frac{d}{dt}I(R;E|Q)_t<0$~\cite{Huang21}. 

Finally, the Breuer-Laine-Piilo (BLP) measure adopts an operational perspective based on state distinguishability~\cite{BLP2}. It relies on the fact that any completely positive trace-preserving (CPTP) map can only decrease the distinguishability between two evolving states $\rho_1(t)$ and $\rho_2(t)$, quantified by the trace distance 
$D(\rho_1(t),\rho_2(t))=\frac{1}{2}|\rho_1(t)-\rho_2(t)|_1$. Hence, any temporary increase in $D(\rho_1(t),\rho_2(t))$ is interpreted as a backflow of information from the environment to the system, and signals non-Markovianity.

Although the RHP, LFS/QMI, QCMI, and BLP measures provide complementary criteria for identifying non-Markovian dynamics, they are insensitive to the nature of the underlying memory. In particular, they do not distinguish whether the observed information backflow originates from genuinely quantum correlations or merely from classical correlations redistributed through the environment. This limitation motivates the discussion in the next section.

\subsection{Quantum Non-Markovianity: Genuine vs Non-Genuine}\label{sec:genuine}
A more refined characterization of memory is provided by the process tensor formalism~\cite{Milz19}, which captures temporal correlations in a fully general multi-time framework. A quantum process over multiple time steps is represented by a higher-order object $\Upsilon_{n:0}$, encoding all accessible correlations across different times, and is Markovian if and only if it factorizes as
\begin{align}
\Upsilon_{n:0} = \Lambda_{n:n-1} \otimes \cdots \otimes \Lambda_{1:0},
\end{align}
where each $\Lambda_{k:k-1}$ is the CPTP map governing the evolution
from time step $k-1$ to $k$. The tensor-product structure signifies the complete absence of temporal correlations, while any deviation indicates non-Markovianity~\cite{Garbellini26}. Importantly, non-Markovianity alone does not reveal the nature of the underlying memory. A process possesses only classical memory if its process tensor $\Upsilon_{n:0}$ admits a convex decomposition into classical mixtures of one-step maps, $\Upsilon_{n:0}=\sum_\lambda p_\lambda
\bigotimes_{k=0}^{n-1}\Upsilon_{k+1:k}^{(\lambda)}$. If no such decomposition exists, the process tensor is entangled across time, and the process exhibits genuine quantum memory~\cite{Giarmatzi2021}. This distinction cannot, in general, be inferred from conventional two-time diagnostics: processes with temporal entanglement may still generate CP-divisible dynamics~\cite{gangwar2026}, whereas purely classical memory can produce non-divisible evolution accompanied by revivals of distinguishability. Notably, non-Markovian features observed in a quantum evolution do not, by themselves, imply a genuinely quantum origin, because the same behavior can also arise from purely classical correlations flowing from the environment back to the system. Measures such as squashed quantum non-Markovianity~\cite{Gangwar_2025}, the local disclosure method~\cite{Charlotte_2024,Charlotte_2026}, non-causal information revival~\cite{Buscemi_2025}, and hysteretic squashed entanglement~\cite{das2026}, among others, have been developed with exactly this separation in mind. We adopt one such state-based diagnostic below.

As noted above, this genuine and non-genuine split is not confined to dynamical maps or process tensors; it can equally be posed at the level of quantum states. This is often the more convenient route in practice: rather than tracking revivals in distinguishability or violations of CP divisibility over time, one can ask directly which part of the observed correlations is intrinsically quantum and which part can be removed by classical conditioning.

Consider a tripartite quantum state $\rho_{ABC}$. A standard quantity used to diagnose non-Markovian correlations is the quantum conditional mutual information (QCMI), $I(A;C \mid B)_\rho = S(\rho_{AB}) + S(\rho_{BC}) - S(\rho_B) - S(\rho_{ABC})$. While $I(A;C \mid B)_\rho = 0$ characterizes quantum Markov states, a strictly positive value does not necessarily imply genuinely quantum non-Markov states. Indeed, classical mixtures of Markov states can also yield $I(A;C \mid B)_\rho > 0$. Thus, QCMI serves as a witness to non-Markovian correlations but does not identify their underlying origin.

To address this limitation, the concept of squashed quantum non-Markovianity (sQNM) has been introduced. This measure removes non-genuine contributions by allowing extensions of the conditioning system and is defined as
\begin{align}
N_{\mathrm{sq}}(A;C \mid B)_\rho := \frac{1}{2} \inf_{\sigma_{ABCE}} I(A;C \mid BE)_\sigma,
\end{align}
where the infimum is taken over all extensions $\sigma_{ABCE}$ such that $\mathrm{Tr}_E[\sigma_{ABCE}] = \rho_{ABC}$. The central idea is that any non-Markovianity that vanishes upon conditioning on an extended system $BE$ cannot be genuinely quantum. Only the component that remains after this optimization represents genuine quantum non-Markovian correlations.

This perspective establishes genuine quantum non-Markovianity as a well-defined quantum resource~\cite{Gangwar_2025}. In particular, the set of states with vanishing sQNM is convex, implying that genuine quantum non-Markovianity cannot be generated by classical mixing alone. A dynamical counterpart of this state-based criterion is provided by the concept of \emph{non-causal information revivals}~\cite{Buscemi_2025}. In this framework, a revival of correlations between a reference and the system is considered non-causal if it disappears upon introducing an inert extension $F$, where $F$ remains completely isolated from the system throughout the evolution. In such cases, the apparent revival can be fully explained by conditioning on the extended environment rather than by invoking a genuine backflow of quantum information. This criterion is the dynamical analog of the sQNM construction and plays a central role in the diagnostic employed in Sec.~\ref{sec:discrete_exampl} and Sec.~\ref{sec:count_example}, where we investigate whether the SWITCH-induced revivals of $I(R;Q)$ survive upon extending the correlations to $I(R;QF)$. 

\section{Analysis of Quantum SWITCH Dynamics through Non-Markovianity}\label{S3}
In this section, we analyze the memory effects induced by the quantum SWITCH, with the aim of clarifying their structural origin and evaluating whether they correspond to genuine quantum non-Markovianity.

We begin by analyzing the constituent dynamics in the absence of the quantum SWITCH. Specifically, we examine the evolution of the mutual information for each individual process and, whenever non-Markovian behavior is observed, we determine whether it is genuinely quantum using the non-causal information revival framework. We then turn to the effective post-selected dynamics generated by the quantum SWITCH and analyze the corresponding mutual information. Whenever revivals of mutual information occur, we again analyze the dynamics with extension to ascertain whether the observed memory effects originate from genuine quantum memory or are instead attributable to classical correlations.

We employ the above formalism to present our main results. In particular, we investigate a large class of both discrete and continuous dynamical processes that exhibit non-Markovian behavior under the action of the quantum SWITCH. We show that despite observed information revivals, these processes do not generate genuine quantum non-Markovianity, even after applying the quantum SWITCH.

\subsection{Example: Discrete-Time Dynamics}\label{sec:discrete_exampl}
\begin{figure}[h]
\centering
\includegraphics[width=0.45\textwidth]{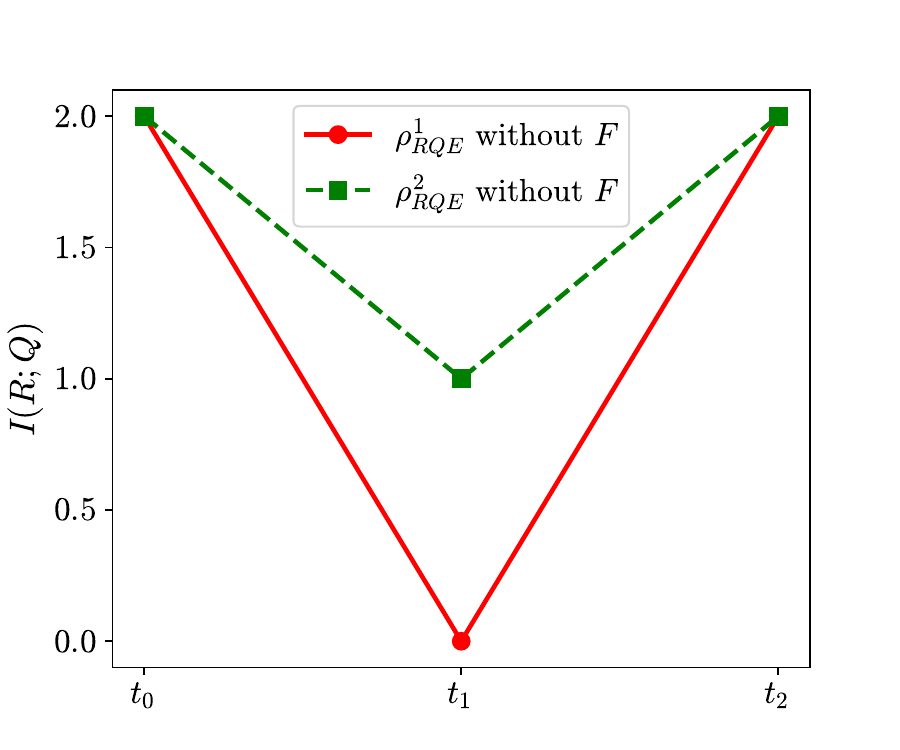}
\caption{\justifying Mutual information $I(R;Q)$ between the reference system ($R$) and the quantum system ($Q$) at discrete time steps $(t_0, t_1, t_2)$, exhibiting non-Markovian dynamics, where the environment is initialized in  $\gamma^{(1)}_E$ maximally mixed state and  $\gamma^{(2)}_E$ classically correlated state.}
\label{fig:a}
\end{figure}

\begin{figure}[h]
\centering
\includegraphics[width=0.5\textwidth]{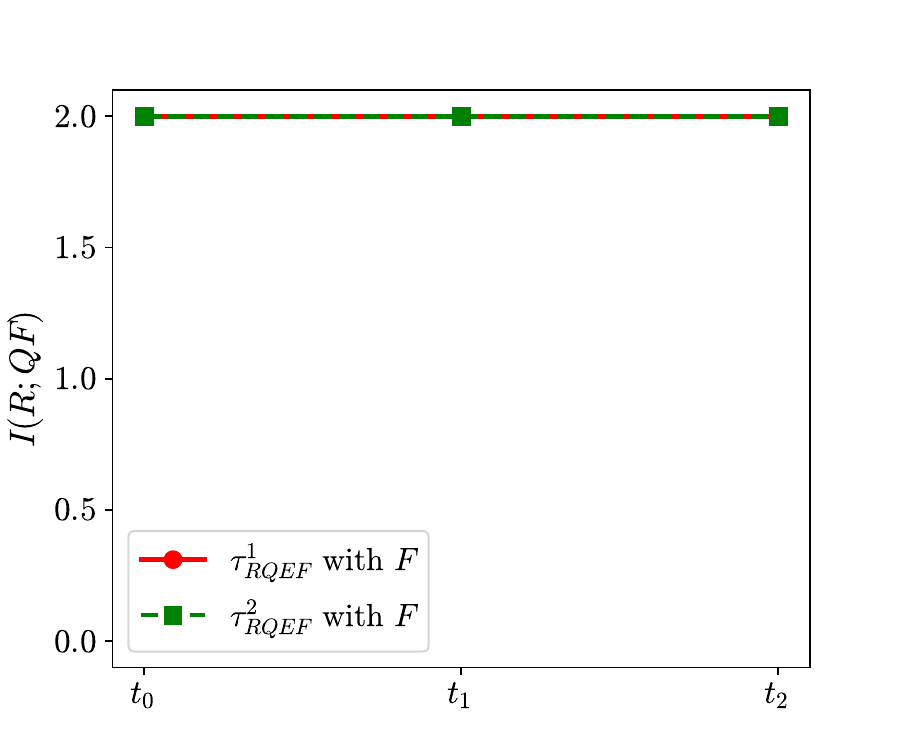}
\caption{\justifying Mutual information $I(R;QF)$, where $F$ is the extended space, shows monotonic behavior at all time steps, confirming the absence of genuine quantum non-Markovianity for both processes.}
\label{fig:extens_indivudal_plot}
\end{figure}

We begin with examples by considering the following initial states (i.e., at time $t_0$) for Process 1 and Process 2, respectively:
\begin{align}
&\rho^{1}_{0} = \ket{\Phi^+}_{RQ}\bra{\Phi^+} \otimes \gamma_E^{(1)}, \\
&\rho^{2}_{0} = \ket{\Phi^+}_{RQ}\bra{\Phi^+} \otimes \gamma_E^{(2)},
\end{align}
where $\ket{\Phi^+}_{RQ}$ is the maximally entangled states shared between the reference ($R$) and the quantum system ($Q$), $\gamma_E^{(1)} = \mathbb{I}/4$ is the maximally mixed state of the environment, and $\gamma_E^{(2)} = \frac{1}{2}(\proj{00}+\proj{11})_E$ is a classically correlated environmental state. The unitary interactions are
\begin{align}
&U^{1}_{QE} = \sum_{i=0}^{3} \sigma^i_{Q} \otimes \ketbra{i}_E, \\
& U^{2}_{QE} = \sum _{i=0}^1 \sigma^i_Q \otimes (\proj{i}\otimes \mathbb{I})_E,
\end{align}
where the reference system $R$ undergoes an identity channel, $\{\sigma^i\}_{i=0}^3$ are the Pauli operators with $\sigma^0 = \mathbb{I}_{2}$, and $\{\ket{i}\}$ forms an orthonormal basis of the environment space. The state at time $t_1$ for both processes is given by
\begin{align}
\rho^j_1 = U^{(j)}_{QE}\, \rho^j_0 \, U^{(j)\dagger}_{QE}, \quad \quad j=\{1,2\},
\end{align} 
and its subsequent evolution to time $t_2$ (where $0 < t_1 < t_2$) yields
\begin{align}
\rho^j_2 = U^{(j)}_{QE}\, \rho^j_1 \, U^{(j)\dagger}_{QE}, \quad \quad j=\{1,2\}.
\end{align}
An analysis of the quantum mutual information $I(R;Q)$ across these time steps reveals non-monotonic behavior for both processes (see Fig.~\ref{fig:a}), indicating the presence of non-Markovian dynamics. However, when the environment is extended for each process individually, the corresponding mutual information $I(R;QF)$ exhibits a monotonic decrease, as shown in Fig.~\ref{fig:extens_indivudal_plot}. Specifically, we extend the initial states as
\begin{align}
    \rho^1 \Rightarrow \tau^1 = \ket{\Phi^+}\bra{\Phi^+}_{RQ} \otimes \ket{\Psi}\bra{\Psi}_{EF_1}, \\ 
    \rho^2 \Rightarrow \tau^2 = \ket{\Phi^+}\bra{\Phi^+}_{RQ} \otimes \ket{\Psi}\bra{\Psi}_{EF_2},
\end{align}
where
\begin{align}
    &\ket{\Psi}_{EF_1} = \frac{1}{2}\sum_{i=0}^{3} \ket{i}_E \otimes \ket{i}_{F_1}\\
    &\ket{\Psi}_{EF_2} = \frac{1}{\sqrt{2}}\left(\ket{00}_E\otimes\ket{0}_{F_2} + \ket{11}_E\otimes\ket{1}_{F_2}\right).
\end{align}
This demonstrates that one can always construct an extended initial state together with a corresponding extended unitary evolution such that the apparent (non-causal) revival of correlations between the reference and the system disappears once the inert extension $F$ is taken into account, provided that no dynamics act on $F$ throughout the evolution.

According to the criterion introduced in Ref.~\cite{Buscemi_2025}, neither of the constituent processes exhibits genuine quantum non-Markovianity. The non-monotonic behavior observed in Fig.~\ref{fig:a} therefore arises entirely from classical, non-causal revivals  of correlations mediated through the environment. This naturally motivates examining whether placing these processes in a coherent superposition of causal orders through the quantum SWITCH can induce genuinely quantum non-Markovian dynamics. We address this question in the following.

\begin{figure}[h]
    \centering
\includegraphics[width=0.45\textwidth]{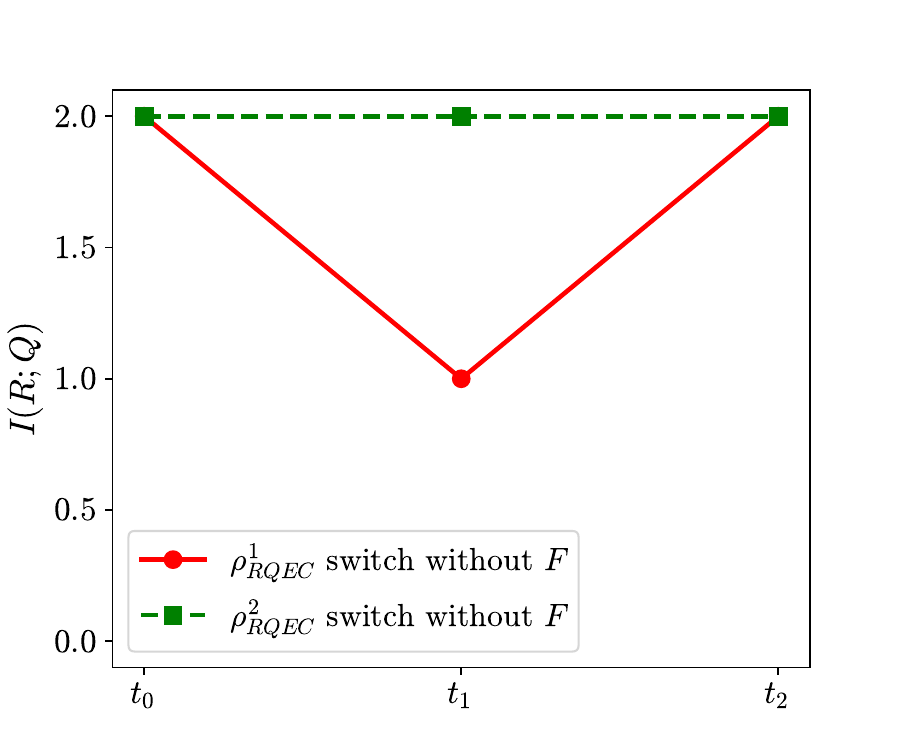}
\caption{\justifying Mutual information $I(R;Q)$ under the quantum SWITCH, where initial state $\rho^1_0$ exhibits non-monotonic (non-Markovian) behavior while initial state $\rho^2_0$ remains constant across all time steps.}
\label{fig:withou_exten_QS}
\end{figure}

The quantum SWITCH applies process $U$ before $V$ if the control qubit is prepared in state $\ket{0}_C$, and $V$ before $U$ if the control qubit is prepared in $\ket{1}_C$, as discussed in Sec.~\ref{subsec:QS}. With the control initialized in the superposition state
\begin{align}
\ket{+}_C = \frac{1}{\sqrt{2}}\left(\ket{0}_C + \ket{1}_C\right),
\end{align}
both causal orders are coherently and simultaniously activated. The initial states on the full composite system $R \otimes Q \otimes E \otimes C$ are
\begin{align}
&\rho^{1}_0 = \ket{\Phi^+}\bra{\Phi^+}_{RQ} \otimes \gamma_E^{(1)}
\otimes \ket{+}\bra{+}_C, \\
&\rho^{2}_0 = \ket{\Phi^+}\bra{\Phi^+}_{RQ} \otimes \gamma_E^{(2)}
\otimes \ket{+}\bra{+}_C,
\end{align}
and the switch unitary acting on $RQEC$ is defined as
\begin{align}
U_{\text{QS}} = \left(U \cdot V\right) \otimes \ket{0}\bra{0}_C 
+ \left(V \cdot U\right) \otimes \ket{1}\bra{1}_C,
\end{align}
where $U = \mathbb{I}_R\otimes U^{1}_{QE}$ and $V =\mathbb{I}_R\otimes U^{2}_{QE}$ are the unitaries of Process 1 and Process 2, respectively. At time $t_1$
\begin{align}
    \rho_1^{j} = U_{\text{QS}} \  (\rho_0^{j}) \ U_{\text{QS}}^\dagger \qquad j = \{1,2\}
\end{align}
and at time $t_2$
\begin{align}
    \rho_2^{j} = U_{\text{QS}} \  (\rho_1^{j}) \ U_{\text{QS}}^\dagger  \qquad j = \{1,2\}.
\end{align}
After the evolution, we perform a projective measurement on the control qubit. The post-measurement states conditioned on outcome $\ket{+}_C$ are $\rho_{RQEC}^i = M \rho^i M^\dagger$ for $i\in\{1,2\}$, where $M = \proj{+}_C$. The corresponding normalized states after the action of quantum SWITCH are
\begin{align}
\frac{\rho_{RQEC}^i}
{\mathrm{Tr}\!\left[\rho_{RQEC}^i\right]}, 
\quad i = 0, 1,
\end{align}
for each of the initial states $\rho^{1}_0$ and $\rho^{2}_0$. We next evaluate the mutual information $I(R;Q)$ for the normalized post-selected states. As shown in Fig.~\ref{fig:withou_exten_QS}, the dynamics corresponding to $\rho^{1}_0$ exhibits a revival,
\begin{align}
I(R;Q)_{t_0} \geq I(R;Q)_{t_1} \leq I(R;Q)_{t_2},
\end{align}
indicating non-Markovian behavior. In contrast, the mutual information associated with $\rho^{2}_0$ remains constant throughout the evolution (see Fig.~\ref{fig:withou_exten_QS}),

\begin{figure}[h]
    \centering    \includegraphics[width=0.5\textwidth]{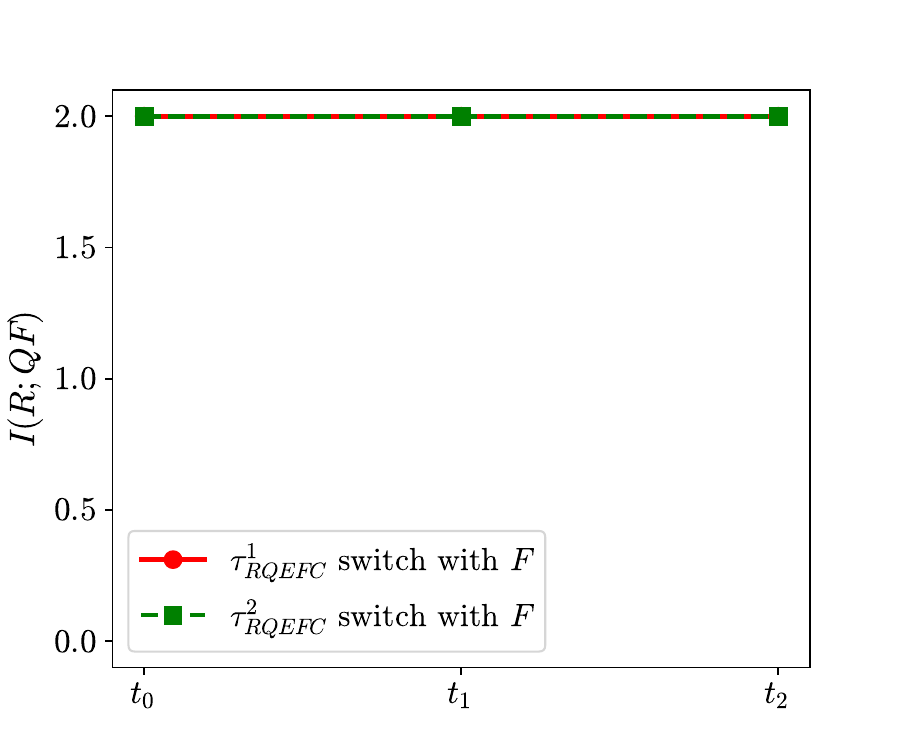}
    \caption{\justifying Mutual information $I(R;QF)$ at discrete time steps $(t_0, t_1, t_2)$ under the quantum SWITCH, remains monotonic, confirming the absence of genuine quantum non-Markovianity.}
    \label{fig:with_exten_QS}
\end{figure}

To determine whether the observed revival is genuinely quantum, we repeat the analysis by extending the environment. The quantum SWITCH now acts on the extended initial states
\begin{align}
\rho^{1}_0 = \ket{\Phi^+}\bra{\Phi^+}_{RQ} \otimes \ket{\Psi}\bra{\Psi}_{EF_1}\otimes \ket{+}\bra{+}_C, \\ 
\rho^{2}_0 = \ket{\Phi^+}\bra{\Phi^+}_{RQ} \otimes \ket{\Psi}\bra{\Psi}_{EF_2}\otimes \ket{+}\bra{+}_C,
\end{align}
where the auxiliary system ($F$) undergoes trivial (identity) evolution. The resulting mutual information $I(R;QF)$ is monotonic for both processes, with no information revival (see Fig.~\ref{fig:with_exten_QS}).
According to the criterion of Ref.~\cite{Buscemi_2025}, the monotonicity of $I(R;QF)$ establishes that the observed information revivals are not genuinely quantum. For the individual processes, although $I(R;Q)$ exhibits non-monotonic behavior (Fig.~\ref{fig:a}), the extended mutual information $I(R;QF)$ becomes monotonic after introducing an appropriate environmental extension (Fig.~\ref{fig:extens_indivudal_plot}), demonstrating that the apparent memory effects are non-genuine. Likewise, while the quantum SWITCH induces a revival of $I(R;Q)$ for $\rho^1_0$ (Fig.~\ref{fig:withou_exten_QS}), the corresponding extended mutual information $I(R;QF)$ remains monotonic (Fig.~\ref{fig:with_exten_QS}). Therefore, although the quantum SWITCH can revive system-reference correlations, it does not generate genuine quantum non-Markovianity; the observed correlation revivals arise solely from an incomplete description of the environment.

\subsection{Example: Continuous Dynamical maps}\label{sec:count_example}

\begin{figure}[h]
\centering
\includegraphics[width=0.45\textwidth]{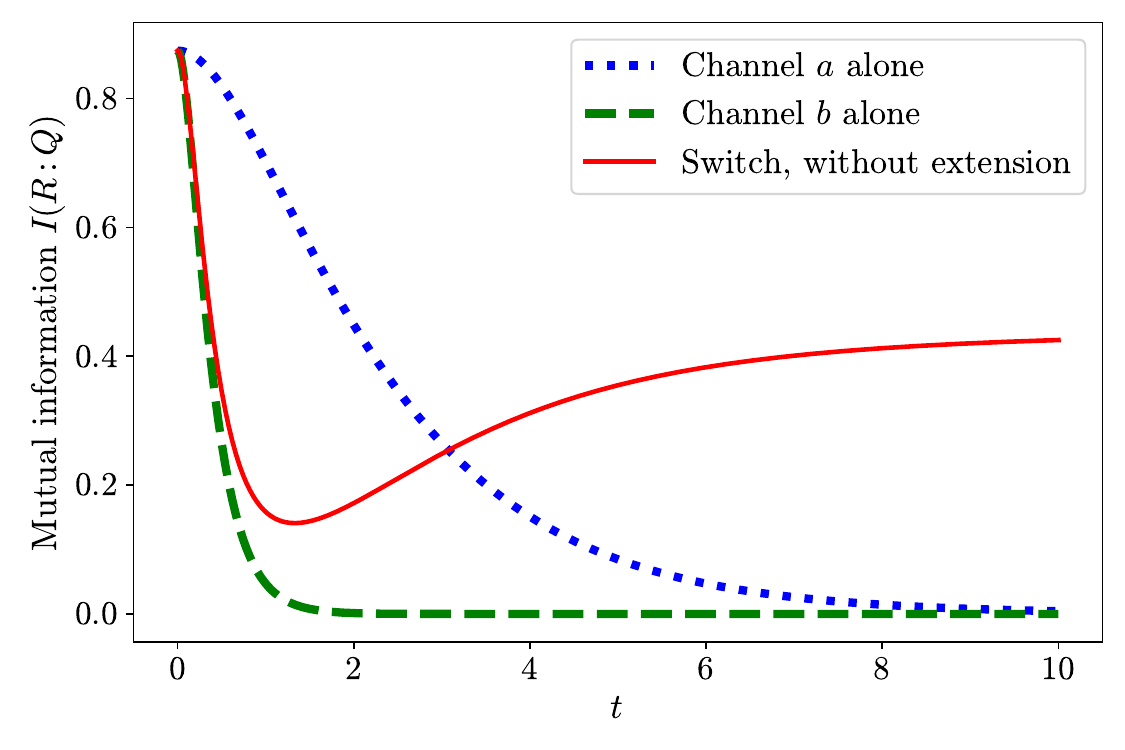}
\caption{\justifying Mutual information $I(R;Q)(t)$ for unitary channel $a$ and unitary channel $b$ alone, both decaying monotonically, but SWITCH dynamics after post-selecting $C$ on $\ket{+}_C$ shows non-monotonic behavior.}
\label{fig:QS_cont_noext}
\end{figure}

We now extend our analysis to a continuous-time setting to investigate whether the conclusions established in Sec.~\ref{sec:discrete_exampl} for discrete unitary interactions continue to hold. We consider a reference system $R$ and a target system $Q$, initially prepared in the Werner state 
\begin{align}
\rho_{RQ}(0) = p\, \proj{\Phi^+}_{RQ} + \frac{1-p}{4}\, \mathbb{I}_R \otimes \mathbb{I}_Q,
\end{align}
(where, for illustration, we set $p=0.7$ for the plots). Unlike the pure Bell state considered in Sec.~\ref{sec:discrete_exampl}, the initial state is mixed. The system $Q$ interacts with an environment ($E$), initially prepared in the maximally mixed state $\gamma_E = {\mathds{1}_E}/{2}$, through the combined unitary
\begin{align}
M(\theta) =
\begin{pmatrix}
1 & 0 & 0 & 0 \\
0 & \cos\theta & \sin\theta & 0 \\
0 & -\sin\theta & \cos\theta & 0 \\
0 & 0 & 0 & 1
\end{pmatrix}_{QE},
\end{align}
which provides a Stinespring dilation of the amplitude-damping channel, transferring a population $|\sin^2\theta|$ from $Q$ to $E$. We then define the two continuous-time unitary evolutions
\begin{align}
U_a(t) &= \mathbb{I}_R \otimes M\!\left(\theta_a(t)\right), \\
U_b(t) &= \mathbb{I}_R \otimes \Big(H_Q\, M\!\left(\theta_b(t)\right)\, H_Q\Big),
\end{align}
where $H_Q$ denotes the Hadamard gate acting on $Q$. Thus the channel $U_b$ realizes amplitude damping in the $\{\ket{+},\ket{-}\}$ basis rather than the computational basis ($\{\ket{0},\ket{1}\}$). The time-dependent parameters are chosen as
\begin{align}
\theta_a(t) = \frac{\pi}{2}\Big(1-e^{-\gamma_a t}\Big), \qquad
\theta_b(t) = \frac{\pi}{2}\Big(1-e^{-\gamma_b t}\Big),
\end{align}
with $\gamma_a = 0.3$ and $\gamma_b = 1.8$. Since both angles increase monotonically from $0$ to $\pi/2$ with time, each channel on its own is CP-divisible: population flows only from the system ($Q$) to the environment ($E$), with no backflow. Consequently, both channels are Markovian as confirmed by the monotonic decay of the mutual information ($I(R;Q)(t)$) shown in Fig.~\ref{fig:QS_cont_noext}. 

We next combine these channels through a quantum SWITCH, with the control qubit $C$ prepared in $\ket{+}_C$. The corresponding switch unitary is
\begin{align}
W(t) = U_b(t)\,U_a(t) \otimes \ket{0}\bra{0}_C + U_a(t)\,U_b(t) \otimes \ket{1}\bra{1}_C,
\end{align}
that acts on $\rho_{RQ}(0)\otimes \gamma_E \otimes \ket{+}\bra{+}_C$. Following the evolution, the control qubit is projectively measured in the $\{\ket{+}\bra{+}, \ket{-}\bra{-}\}$ basis. Conditioned on the measurement outcome, the resulting post-selected state is normalized, and the corresponding mutual information $I(R;Q)(t)$ is subsequently evaluated. In contrast to the individual channels, the SWITCH-induced dynamics exhibits a clear non-monotonic evolution of $I(R;Q)(t)$ as shown in Fig.~\ref{fig:QS_cont_noext}. This phenomenon, referred to as emergent non-Markovianity~\cite {Anand_25}, is characterized by intervals for which $\frac{d}{dt}I(R;Q)_t>0$, thereby satisfying the LFS criterion for quantum non-Markovianity~\cite{Luo2012}, despite each constituent channel is individually Markovian.

\begin{figure}[h]
\centering
\includegraphics[width=0.45\textwidth]{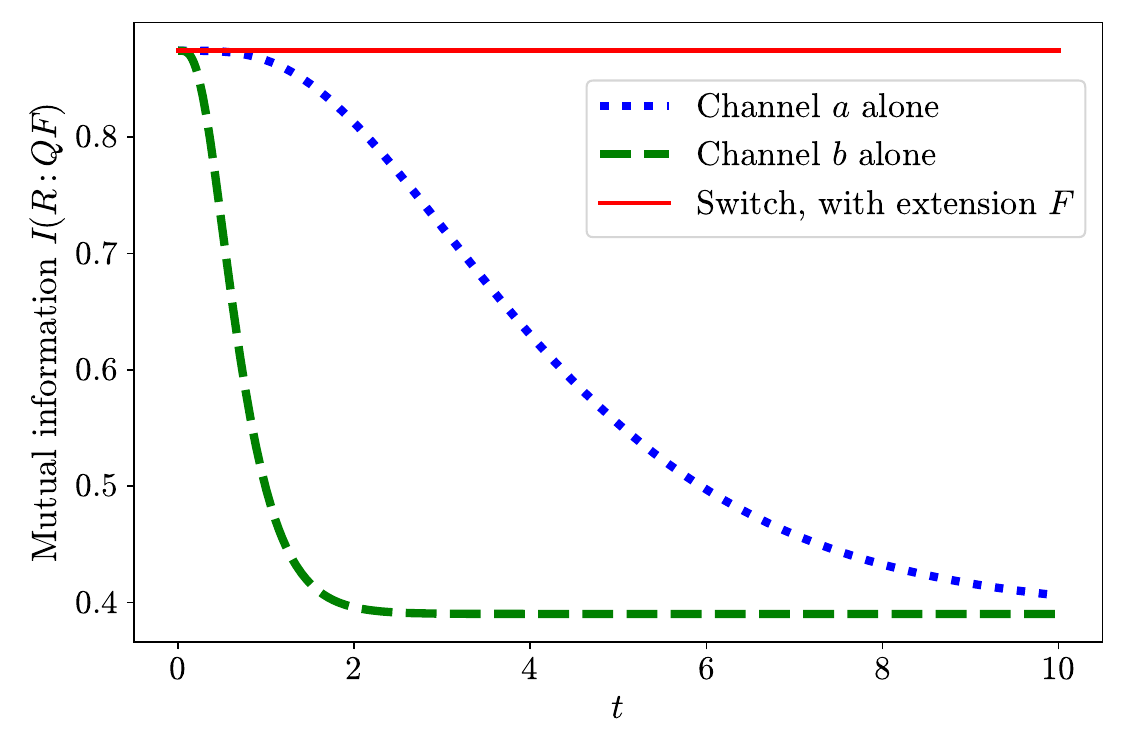}
\caption{\justifying Mutual information $I(R;QF)(t)$, where $F$ is an inert extension. Both individual channels remain monotonic, while the switch-induced revival is entirely absent: $I(R;QF)(t)$ stays essentially constant at all times, confirming the absence of genuine quantum non-Markovianity.}
\label{fig:QS_cont_ext}
\end{figure}

Following the approach adopted in Sec.~\ref{sec:discrete_exampl}, we next extend the environment by introducing an inert extension $F$, which remains dynamically intact throughout the evolution. Instead of preparing the environment $E$ in the maximally mixed state $\gamma_E$, we consider the (extended) purified initial state $\ket{\Phi^+}_{EF}$, satisfying $\mathrm{Tr}_F\big[\ket{\Phi^+}_{EF}\bra{\Phi^+}\big] = \gamma_E$, with $F$ left completely idle throughout the dynamics. The unitary channels, as well as the quantum SWITCH, continue to act exclusively on the subsystem $(Q,E)$, while both the reference system $R$ and the extension $F$ evolve trivially under the identity channel. We then evaluate the extended mutual information $I(R;QF)(t)$ instead of $I(R;Q)(t)$. 
For the individual channels, the mutual information $I(R;QF)(t)$ remains monotonic (Fig.~\ref{fig:QS_cont_ext}). Thus, the apparent information backflow observed in $I(R;Q)(t)$ under the quantum SWITCH disappears upon extension.
The restoration of monotonicity in $I(R;QF)(t)$ demonstrates that the revival observed in $I(R;Q)(t)$ under the quantum SWITCH (Fig.~\ref{fig:QS_cont_noext}) does not correspond to genuine quantum non-Markovianity. Therefore, even in the continuous-time setting, quantum SWITCH induced non-Markovianity is not genuinely quantum.

We note a structural distinction from the discrete-time example of Sec.~\ref{sec:discrete_exampl}. In that case, the initial state of the system-reference pair ($RQ$) was pure, so purifying the environment $E$ resulted in an overall pure state of the extended system. By contrast, in Sec.~\ref{sec:count_example}, the initial state $\rho_{RQ}(0)$ is a mixed Werner state and remains mixed even after extending $E$ via $F$. The restoration of monotonicity in $I(R;QF)(t)$ nonetheless persists, demonstrating that the environmental extension criterion of Ref.~\cite{Buscemi_2025} does not rely on the purity of the system-reference state and applies equally to mixed-state scenarios.

\section{Discussions}\label{S4}

Indefinite causal order, with the quantum SWITCH as its canonical realization, has been widely regarded as a useful resource for achieving advantages across diverse quantum tasks, including communication, computation, and thermodynamics~\cite{Ebler_18,Chiribella_21,Bhattacharya_21,Araujo14,Guerin16,Zhao_20,Tamal_20,Vedral_20,Mukhopadhyay19,Liu_23}. Despite these developments, the fundamental origin of such advantages remains unsettled. In particular, it is still unclear whether they stem intrinsically from indefinite causal structure or can instead be reproduced by suitable superpositions of fixed-order processes \cite{Guerin19}. Recent studies have identified several operational ingredients underlying these advantages, including coherence in the initial control system and in the measurement basis~\cite{Anand_25}, as well as the Hilbert-space dimension of the control system~\cite{Mukherjee24}. Complementary to these viewpoints, from a dynamical perspective, non-Markovian memory effects have also been proposed as a potential resource associated with the quantum SWITCH~\cite{Maity_24,Anand_25}.

Motivated by this perspective, here we have investigated the nature of the memory generated by the quantum SWITCH, with particular emphasis on distinguishing genuinely quantum non-Markovianity from effects attributable to classical correlations or statistical mixing. Using both discrete-time models and continuous-time dynamical maps, we have analyzed the resulting dynamics under the quantum SWITCH. Our results demonstrate that, although signatures of non-Markovianity may appear according to the standard criteria, these effects do not persist under appropriate extensions of the environment and hence do not qualify as genuinely quantum non-Markovian. Consequently, our results call for a more careful reassessment of the physical resources underlying the reported quantum advantages of indefinite causal order and motivate a systematic search for the genuine quantum resources responsible for these advantages beyond non-Markovian memory effects.\\

{\it Acknowledgment:} R.G. acknowledges the support from the Israel Science Foundation under Grant No. 1192/24, and AGM acknowledges the financial support provided by the IIT Goa Startup Grant (No. 2025/SG/AGM/058).
\bibliography{main}

\appendix

\end{document}